\documentclass[twocolumn,10pt,a4paper,empty]{article}
\usepackage{graphicx}
\begin{document}

\pagestyle{empty}

\title{\bf Emergence of 
scale invariance and efficiency \\ in a racetrack betting market}

\author{S.~Mori $^{\rm *}$ and M.~Hisakado $^{\dag}$
\\
\\
Department of Physics, School of Science, Kitasato University
\\
1-15-1 Kitasato, Sagamihara 228-8555, Japan
\\
\\
Standard and Poor's
\\
1-6-5 Marunouchi, Chiyoda-ku, Tokyo 100-0005, Japan
} %

\date{KEY WORDS: scale invariance, efficiency, racetrack betting market,
emergence}

\maketitle\thispagestyle{empty} 

\begin{abstract}
 We study the time change of the relation between the rank 
 of a racehorse in the Japan Racing Association and the result of victory 
 or defeat. Horses are ranked according to the win bet fractions. As the
 vote 
 progresses, the racehorses are mixed on the win bet fraction axis.
 We see the emergence of a scale invariant relation  between the cumulative 
 distribution function of the winning horse $x_{1}$ and that  of 
 the losing horse $x_{0}$. $x_{1}\propto x_{0}^{\alpha}$ holds 
 in the small win bet fraction region. We also see the efficiency of the 
 market as the vote proceeds. However, the convergence to the efficient state 
 is not monotonic. The time change of the distribution of a vote 
is complicated.
 Votes resume concentration on popular horses, after the distribution
 spreads to a certain extent. 
 In order to explain scale invariance, we introduce a simple 
 voting model. In a `double' scaling  limit, we show that the exact scale 
 invariance relation $x_{1}=x_{0}^{\alpha}$ holds over the entire range 
 $0\le x_{0},x_{1}\le 1$.
\end{abstract}

\vspace*{0.6cm}

\section{Introduction}
 Racetrack betting is a simple exercise of gaining a profit or losing
 one's wager. However, 
 one needs to make a decision in the face of  uncertainty
 and a closer inspection reveals great complexity and scope. 
 The field has attracted many academics from a wide 
 variety of  disciplines and has become a subject of wider 
 importance \cite{Ziemba}. 
 Compared to the stock or currency 
 exchange markets, racetrack betting is a short-lived and repeated market.
 It is possible to obtain starker views of aggregated better behaviour
 and study the market efficiency. 
 One of the main findings of the previous
 studies is the `favorite-longshot bias' in the racetrack betting market 
 \cite{Griffith,Ziemba2}. Final odds are, on average, accurate measures
 of winning and 
 short-odds horses are systematically undervalued and long-odds horses are 
 systematically overvalued.
 
 From an econophysical viewpoint, racetrack
 betting is an interesting subject. 
 Park and Dommany have done 
 an analysis of the distribution of final odds (dividends) of 
 the races organized by
 the Korean Racing Association \cite{Park}. 
 They found power law behaviour in the distribution and proposed a 
 simple betting model. 
 Ichinomiya also found the power law 
 in the races of the Japan Racing Association 
(JRA) \cite{Ichinomiya} and proposed
 another betting model. 
 We have studied the relation between 
 the rank of a racehorse in JRA and the result of victory 
 or defeat \cite{Mori}. Horses are ranked according to the win bet fractions. 
 We studied the 
 distribution of the winning horses in the long-odds region \cite{Mori}. 
 Between the cumulative distribution function of the winning horses  
 $x_{1}$ and that of the losing horses $x_{0}$, we find a scale
 invariant relation $x_{1}\propto x_{0}^{\alpha}$ with $\alpha=1.81$.
 We show that in a `P\'{o}lya' like betting model, where
 betters vote on the horses according to  the probabilities that are
 proportional to the votes, such a scale invariance emerges 
 in a self-organized fashion. 
 Furthermore, the exact scale invariant
 relation $x_{1}=x_{0}^{\alpha}$ holds exactly over the entire range 
$0\le x_{0},x_{1}\le 1$ in a limit. 
 We also studied a voting model with two kinds
 of voters, independent and copycat \cite{Hisakado}. We find 
 that a phase transition occurs in the process of information cascade and 
 it causes a critical slowing down in the convergence of the decision making of 
 crowds. 
 
 In this paper, we study the time series data of the vote¡¡in 
 JRA races in detail. We show that as the vote progresses, scale
 invariance and market efficiency do emerge in the rank of the racehorses.  
 The decision making process in betting is not simple and some  subtle 
 mechanism does work. We explain the scale invariance based on a simple
 betting model.
    
\section{Racetrack betting process}
 
 We study the win bet 
data of JRA races in 2008. A win bet requires one to name the winner of
the race. There are 3542 races and 
 we choose  3250 races whose final public win pool (total number of votes) 
$V_{r}$ is in the range
 $10^{5}\le V_{r} \le 10^{6}, r \in \{1,2,\cdots,R=3250 \}$. 
 $H_{r}$ horses run in race $r$ and $7 \le H_{r} \le 18$.
We remove 133 cancelled horses and the total number of horses  
$N\equiv \sum_{r=1}^{R}H_r$ is $47273$. There are 3251 winning horses
 (one tie occurs) and are denoted as $N_{1}=3251$.
Of course, the remaining 44022 horses are losers and are denoted as 
$N_{0}=N-N_{1}$. 
The number of times race $r$ is announced is
denoted as $K_{r}$ and is
in the range $13\le K_{r} \le 262$. The total number of announcements is
$K\equiv \sum_{r=1}^{R}K_{r}=285269$.
The time to the race entry time (start of the race) in minutes is denoted 
as $T^{r}_{k}$.
 We denote the odds of the $i$th horse in race $r$ at the $k$th announcement  
as $O_{i,k}^{r}$ and the public win pool as $V^{r}_{k}$. 
$I^{r}_{i}$ denotes the results of the races. $I^{r}_{i}=1 (0)$ implies
 that the horse wins (loses). A typical sample from the data is shown
in Table \ref{tab:TS}.

\begin{table}[htbp]
\caption{\label{tab:TS}
Time Series of odds and pool for a race that starts at $13:00$.
$H_{r}=10,K_{r}=52$ and we show the data only for the first three horses
$1\le i \le 3$.
The first horse wins in the race $(I^{r}_{1}=1,I^{r}_{2}=I^{r}_{3}=0)$.}
\begin{tabular}{ccccccc}
k & $T^{r}_{k}$ [min] 
& $V^{r}_{k}$ & $O^{r}_{1,k}$ & $O^{r}_{2,k}$& $O^{r}_{3,k}$ & $\cdots$ \\
\hline
1  & 358 &       1  &    0.0 &  0.0 &    0.0 & $\cdots$  \\  
2 & 351  &      169 &   1.6 &  33.3 &  7.9 &  $\cdots$ \\   
3 & 343  &      314 &   1.8 & 11.3 &  8.0 & $\cdots$ \\
4 & 336  &      812  &  2.9 & 17.8 &  14.6 & $\cdots$  \\
5 & 329  &      1400 &  3.3 &  8.6 & 10.6 & $\cdots$ \\
6 & 322  &      1587 &  2.7 &  9.2 &  11.3 & $\cdots$  \\
$\vdots$ & $\vdots$& $\vdots$& $\vdots$& $\vdots$& $\vdots$ & $\vdots$
 \\
51 & 10  &  80064 &   2.4 &  6.4  & 13.4 & $\cdots$ \\
52 & 4 &  148289 &  2.4 &  4.9 & 16.1 & $\cdots$ \\
53 & -2 &  211653 &  2.4 &  5.3  & 17.0 & $\cdots$   
\end{tabular}
\end{table}

 From $O^{r}_{i,k}$, we estimate the win bet fraction $x^{r}_{i,k}$ by
 the following relation.
\begin{equation}
x^{r}_{i,k}=\frac{0.788}{O^{r}_{i,k}-0.1}.
\end{equation}
If the above does not sum up to one, we renormalize it as 
$\hat{x}^{r}_{i,k}
=\frac{x^{r}_{i,k}}{\sum_{i=1}^{H_{r}}x^{r}_{i,k}}$. Hereafter,
we use $x^{r}_{i,k}$ in place of  $\hat{x}^{r}_{i,k}$ \cite{Votes}.

We use the average public win pool as a time variable $t$ of the
whole betting process.
For each $v$, we choose the nearest $V^{r}_{i,k}$ and
use the average value of $V^{r}_{i,k}$ as the time variable $t$. 
More explicitly, we define $t$ as
\begin{eqnarray}
&&t(v) \equiv \frac{1}{R}\sum_{r=1}^{R} V^{r}_{k^{r}(v)}, \\
&& k^{r}(v)\equiv \{k \hspace*{0.1cm}|\hspace*{0.1cm} 
\mbox{Min}_{k}|V^{r}_{k}-v|  \}.
\end{eqnarray}
We define $x^{r}_{i}(t)$ and $T^{r}(t)$ as
\begin{eqnarray}
x^{r}_{i}(t)&\equiv & x^{r}_{i,k^{r}(v)}  \\
T^{r}(t)&\equiv & T^{r}_{k^{r}(v)}.
\end{eqnarray}
We denote the average value of $T^{r}(t)$ as $T(t)$ and it is defined as 
\begin{equation}
T(t)\equiv \frac{1}{R}\sum_{r=1}^{R}T^{r}_{k^{r}(v)}.
\end{equation}

\begin{figure}[htbp]
\begin{center}
\includegraphics[width=7cm,clip]{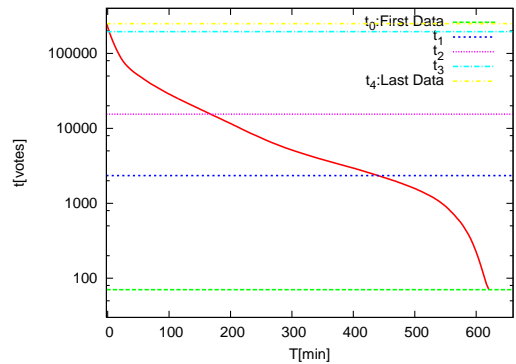}
\end{center}
\caption{Relation between average time to start $T$ 
and average number of votes $t$.
}
\label{fig:Time_vs_Votes}
\end{figure}

Figure \ref{fig:Time_vs_Votes} shows the relation between 
$T$ and $t$.
We see a rapid growth of the average number of
votes $t$ as we approach the start of the race ($T\to 0$). 
Almost half of the votes are thrown 
in the last ten minutes. 
 We choose five timings from the voting process and denote them as
 $t_{j},j\in \{0,1,2,3,4\}$. $t_{0}$ corresponds to the first voting
 data $k=1$ in each race. $t_{0}\simeq 70.4$ and 
$T(t_{0})\simeq 620.6$[min].   
 At $t_{1}\simeq 2342.9$, $T(t_{1})\simeq 440.2$[min]. 
At $t_{2}\simeq 15482.0$,
 $T(t_{2})\simeq 167.6$[min] and at $t_{3}\simeq 195201.3$, 
$T(t_{3})\simeq 2.5$[min].
 $t_{4}$ corresponds to the last voting data $k=K_{r}$ in each race.
 $t_{4}\simeq 249708.8$ and $T(t_{4})\simeq -1.1$[min]. All data are
 summarized after the start of race. The reason why we choose $t_{i},i
 \in \{1,2,3\}$ will become clear in section 4.

In order to see the betting process pictorially, we arrange the 
$N$ horses in the order of the size of $x^{r}_{i}(t)$. We denote the arranged
win bet fraction as $x_{\alpha}(t),\alpha\in \{1,2,\cdots,N\}$.
\begin{equation}
x_{1}(t)\ge x_{2}(t)\ge x_{3}(t)\ge \cdots \ge x_{N}(t)
\end{equation}
$I_{\alpha}(t)$ tells us  whether horse $\alpha$ wins ($I_{\alpha}=1$) or loses
($I_{\alpha}=0$). In general, the probability that the horse with large
$x_{\alpha}(t)$ wins is big and vice versa. 
We arrange the horses in the 
increasing in order of  $\alpha$ from left to right. 
On the left-hand side of the sequence, more strong horses exist. 
On the right-hand side, there are more 
weak horses. If the win bet fraction does not contain any information
about the strength of the horses, $I_{\alpha}(t)$ is one and zero
randomly. Conversely, if the information is perfectly
correct, the first $N_{1}$ horses' $I_{\alpha}(t)$ are 1 and the remaining 
$N_{0}$ horses' $I_{\alpha}(t)$ are 0.  

\begin{figure}[htbp]
\begin{center}
\includegraphics[width=7.0cm,clip]{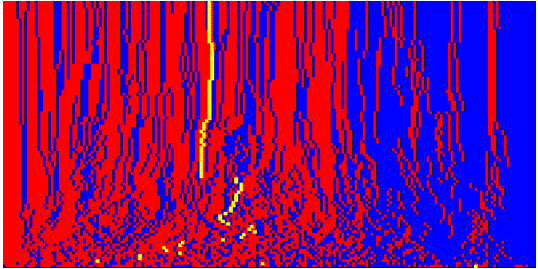}
\end{center}
\caption{Pictorial presentation of the betting process.
We choose 100 winning (red) and 100 losing (blue) horses randomly and follow
their ranking as the betting proceeds. One losing horse is tagged by
 yellow dots.}
\label{fig:Votes_vs_Gradation}
\end{figure}

Figure \ref{fig:Votes_vs_Gradation}
 shows the time change of the ranking process of the horses. 
We choose 100 horses for each category $I_{\alpha}=\{1,0\}$  and follow 
their ranking
as the betting proceeds. We use $\Delta v=2000$ as the time step. 
During one hundred steps, the average number of votes
$t$
changes
from $t_{0}$ for $v=0$ (bottom) to about 178116 for $v=200000$ (top).
$v=0$ corresponds to the first announcement $r=1$ and $t=t_{0}$. 
At $t_{0}$, the horses are arranged in the sequence at random. The win bet 
fractions $x_{\alpha}(t_{0})$
do not contain much information about the strength of the horses.
As the betting progresses, the phase separation between the two categories
of the horses does occur. Winning (losing) horses move to the  left
(right)  in general. In the end, to the left (right) are more winning 
(losing) horses. In the betting process, betters have succeeded in
choosing the winning horses to some extent. We also note that  
there remain winning horses in the right. This means that we can find
 winning horses with very small win bet fractions. 
 
\section{Emergence of scale invariance} 

   In order to see the existence of a winning horse with a very small
   win bet ration or very low rank, we calculate the cumulative 
functions of the winning horses
and that of the losing horses counted from the lowest rank (right). 
More precisely,
we study the following quantities.
\begin{equation}
x_{\mu}(s)=\frac{1}{N_{\mu}}\sum_{\alpha=N(1-s)}^{N}\delta_{I_{\alpha},\mu}
\hspace*{0.2cm},
\hspace*{0.2cm}\mu \in \{0,1\}
\end{equation}
The above  are normalized so that 
$x_{\mu}(s)$ is zero at $s=0$ and 1 at $s=1$.
The curve $(x_{0}(s),x_{1}(s))$ is known as a receiver operating characteristic (ROC) 
curve. We are interested in the limit $s\to 0$. In particular, we 
study the scale invariant relation $x_{1}\propto x_{0}^{\alpha}$.
If such a power law relation holds, we can find winning horses
 with any small win bet fraction.  
 \begin{figure}[t]
(a)
\begin{center}
\includegraphics[width=7.0cm,clip]{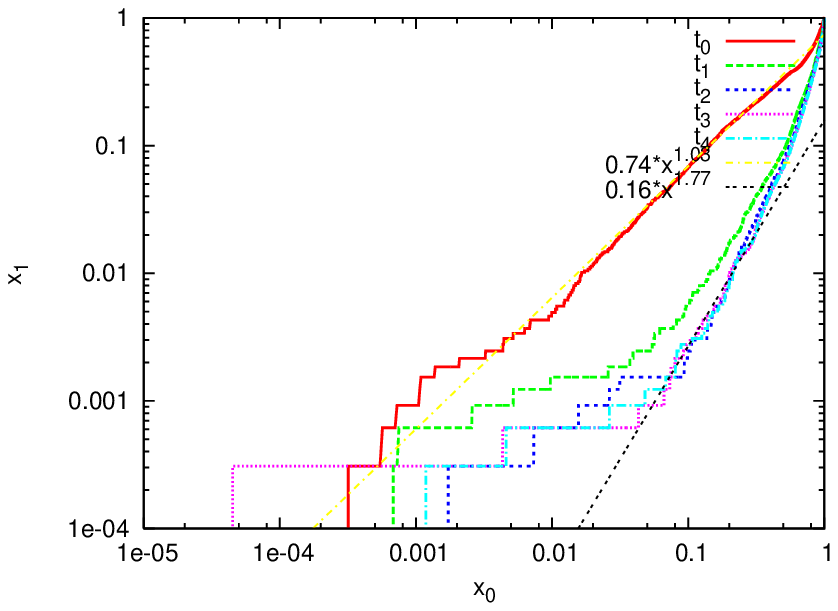}
\end{center}
(b)
\begin{center}
\includegraphics[width=7.0cm,clip]{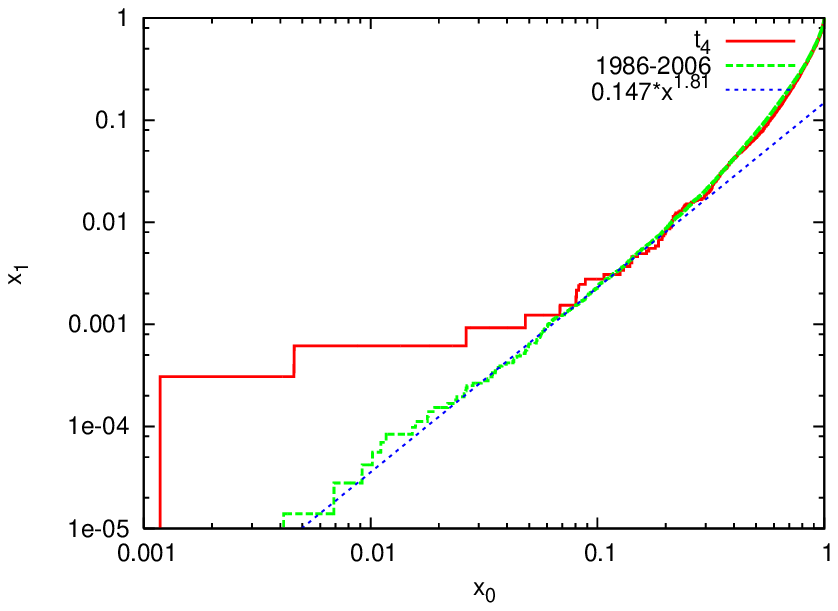}
\end{center}
\caption{
(a) Double-logarithmic plot of the ROC curve $(x_{0},x_{1})$.
We plot four ROC curves for $t=t_{i},i\in \{0,1,2,3,4\}$.
The fitted lines with $x_{1}=a\cdot x_{0}^{\alpha}$ for 
$t_{0}$ and $t_{3}$ are also plotted.
(b) Double-logarithmic plot of the ROC curve $(x_{0},x_{1})$ for 
the data of all 1986-2006 JRA races. 
We also show the fitted line with 
$x_{1}=a\cdot x_{0}^{\alpha}$.}
\label{fig:ROC}
\end{figure}

Figure \ref{fig:ROC}a
 shows the double logarithmic plot of the ROC curves 
 $(x_{0}(s),x_{1}(s))$. We show five curves for $t=t_{i},i\in
 \{0,1,2,3,4\}$.
 At $t=t_{0}$, there are 70 votes in the pool on average. 
The plot is almost a diagonal line
from $(0,0)$ to $(1,1)$, which means that the winning 
and losing horses are mixed 
well in the sequence. 
The curve is fitted with the power relation 
$x_{1}=a\cdot x_{0}^{\alpha}$  with $\alpha=1.03$.  
 As the betting progresses, the curve becomes more downward
 convex shape and the slope increases. At $t=t_{3},t_{4}$, 
the degree of  the phase separation between 
the two types of horses reaches its maximum. On the other hand, we
 also see that there exist winning horses in the very low rank region.  
 Each step upwards of  the curve implies the existence of a 
winning horse and the upward
 movement starts for a very small $x_{0}$. In addition, we also see a
 straight line region in the curve in $0.03\le x_{0}\le 0.3$.
We have fitted the curve as in the previous $(t_{0})$ case and we get $\alpha=1.77$. 
 This means that the  scale invariant relation between $x_{1}$ and
 $x_{0}$ also holds after many rounds of betting.

 In order to see scale invariance more clearly, we show the same plot
 for the data of all JRA races from 1986 to 2006 in Figure \ref{fig:ROC}b
 \cite{Mori}. 
 There are 71549 races and $N_{1}=71650$ and $N_{0}=829716$. 
The plot becomes straight in $0.003\le x_{0} \le 0.3$ 
with $\alpha=1.81$.




\section{Emergence of efficiency}

The win bet fraction aggregates the wisdom of the betters in the racetrack 
betting market. 
In order to quantify the accuracy, we use two
measures. The first one is  the   
accuracy ratio (AR). AR measures how an event occurs 
in the order of a rank. Here, we consider the event that a horse 
wins in the  race. Horses are arranged in the increasing 
order of the size of the win bet
fraction $x_{\alpha}, \alpha\in \{1,2,\cdots N\}$. There are 
$N_{1}$ winning horses and $N_{0}=N-N_{1}$ losing horses. As we have explained
before,
if the prediction of the betters is good, 
the winning horses are
concentrated in the higher ranks. 
If the prediction is perfect, the first
$N_{1}$ horses are the winning ones and $I_{\alpha}$ is one for them.
In order to define the accuracy of the prediction or to measure the 
completeness of the rank, we introduce a Lorenz curve $(x,\mbox{L}(x))$ for
$0\le x, \mbox{L}(x) \le 1$ as
\begin{equation}
\mbox{L}(x)\equiv \frac{1}{N_{1}}\sum_{\alpha=1}^{N\cdot x}
\delta_{I_{\alpha},1}.
\end{equation}

\begin{figure}[h]
(a)
\begin{center}
\includegraphics[width=7.0cm,clip]{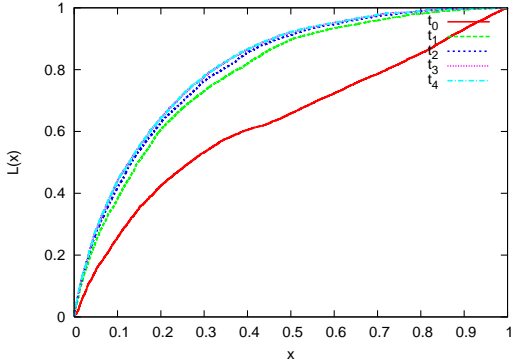}
\end{center}
(b)
\begin{center}
\includegraphics[width=7.0cm,clip]{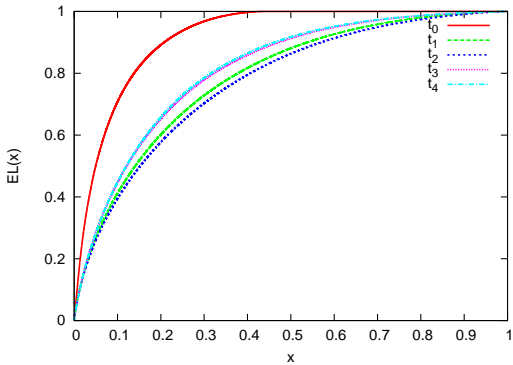}
\end{center}
\caption{
Plot of $(x,$L$(x))$ (a) and $(x,$EL$(x))$ (b) for $t=t_{i},i\in \{0,1,2,3,4\}$.
}
\label{fig:Lorenz}
\end{figure}

Figure \ref{fig:Lorenz}a 
depicts the Lorenz curves for $t=t_{j},j\in {0,1,2,3,4}$.
At $t=t_{0}$, after about seventy rounds, the horses are arranged at random 
on the sequence $\alpha\in \{1,2,\cdots ,N\}$. The Lorenz curve runs almost along a diagonal line.
As the betting proceeds, the winning horses ($I_{\alpha}=1$) 
move to the higher
ranks and the degree of the upward convex nature of the curve
increases. 
The preciseness of the prediction increases monotonically.
At $t=t_{2}$, the increase almost stops and the accuracy of the predictions
reaches a maximum. 

In order to quantify the accuracy of the predictions of the racetrack 
betters, we use accuracy ratio, AR \cite{Enleman}.
AR is defined as
\begin{equation}
\mbox{AR}\equiv
 (\int_{0}^{1}\mbox{L}(x)dx-\frac{1}{2})/\frac{1}{2}(1-\frac{N_{1}}{N})
\label{AR_def}.
\end{equation}
AR measures how different is the ranking from the complete case.
The denominator in the definition is the normalization factor 
that  ensures that AR is one for the completely ordered case.
If the ranking is perfect, $\int_{0}^{1}L(x)dx-\frac{1}{2}=\frac{1}{2}
(1-\frac{N_{1}}{N})$ and AR is 1.

We also introduce an expected Lorenz curve $(x,$EL$(x))$, which is defined
as
\begin{equation}
\mbox{EL}(x)\equiv \frac{1}{R}\sum_{\alpha=1}^{N\cdot x}
x_{\alpha}
\end{equation}
\cite{EL}.
In order to quantify how the bets are concentrated or scattered 
among horse, we introduce expected AR and call it EAR.
EAR is defined as
\begin{equation}
\mbox{EAR}\equiv
 (\int_{0}^{1}\mbox{EL}(x)dx-\frac{1}{2})/\frac{1}{2}(1-\frac{R}{N}).
\end{equation}
If we assume that the horses are divided into two groups, A and B.
The horses' win bet fraction in group A ($x_{\alpha},\alpha\in A$) 
is large and those of 
other horses in group B ($x_{\beta},\beta \in B$) is
small, the votes are concentrated  on the horses in group A. 
EAR is large. If the votes are scattered among all horses, 
EAR is small. In particular, if $x_{\alpha}=\frac{R}{N}$
 for all $\alpha \in N$, EAR is zero. If 
$x_{\alpha}=1,\alpha \in A$ and $x_{\beta}=0,\beta\in B$, 
EAR is one. 

Figure \ref{fig:Lorenz}b
depicts EL$(x)$ for $t=t_{j},j\in {0,1,2,3,4}$.
At $t=t_{0}$, EL$(x)$ rapidly increases to one, which implies that
 votes are concentrated on small number of horses. However, AR is small 
at $t_{0}$ and the horses are not the winning horses. After $t_{0}$, the
votes are scattered among many horses and 
the degree of the upward convex nature of the curve
decreases up to $t_{2}$. After the decline, it begins to increases.
At $t_{4}$, the degree of the concentration of  
votes reaches a maximum. $t_{2}$ is the boundary line.
Before $t_{2}$, votes are more and more distributed among many horses.
After $t_{2}$, the votes tend to be  concentrated on  popular horses.

\begin{figure}[htbp]
\begin{center}
\includegraphics[width=7.0cm,clip]{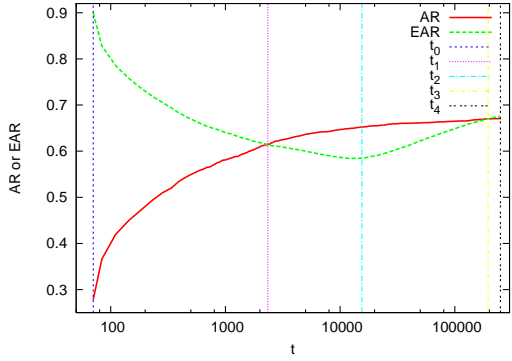}
\end{center}
\caption{Plot of AR and EAR as the functions of $t$. 
The plot starts at $t_{0}$ and the two curves intersects at $t_{1}$ and
 $t_{3}$. At $t_{2}$, the discrepancy reaches its maximum after
 $t_{1}$.}
\label{fig:t_vs_AREAR}
\end{figure}

By comparing the behaviour of L$(x)$ and EL$(x)$, we are able to study 
the efficiency of the market.
If the probability that the horse $\alpha$ wins is $x_{\alpha}$,
the two Lorenz curves L$(x)$ and EL$(x)$ 
do coincide with each other 
in the limit $N \to \infty$. Hence the equality  $\mbox{AR}=\mbox{EAR}$
is a necessary condition of the efficiency of the market.
However, it is not a sufficient condition.
Even if the equality holds, there is a possibility that the
two curves depart from each other.
We also note that if the strength of the horse at rank $x$ is
overvalued, the inequality $\mbox{L}'(x)<\mbox{EL}'(x)$ holds.
On the contrary if the strength is undervalued, 
the inequality $\mbox{L}'(x)>\mbox{EL}'(x)$ holds.
Here $\mbox{L}'(x)=\frac{d \mbox{L}(x)}{dx}$ and 
$\mbox{EL}'(x)=\frac{d \mbox{EL}(x)}{dx}$.

Figure \ref{fig:t_vs_AREAR} shows AR and EAR as the functions of $t$. 
As the betting progresses, AR increases monotonically and it 
almost reaches  its maximum at $t=t_{2}$. Afterwards, the increase in AR
is very slow and the following bets do not increase the accuracy 
of the prediction as to  which horse wins the race.
More interesting behaviour can be found in the time change of EAR.
At $t=t_{0}$, EAR is very large and is nearly 0.9.
Almost all votes are concentrated on small number of horses.
However, AR at $t=t_{0}$ is small and the horses 
with large bet fractions are not so strong.
The true strong horses are scattered all over 
the rank of the win bet fraction. 
Afterwords, EAR decreases rapidly and at $t=t_{1}$, AR
and EAR coincide. 
The necessary condition of the market efficiency is
satisfied at $t=t_{1}$. 
Up to $t=t_{2}$, EAR decreases and almost  reaches  
its minimum. The bets are scattered among many horses and this
implies the rich
variety of the betters' predictions as to which horse wins the race. 
This also means that the strong horses are undervalued and the weak horses 
are overvalued, that is the `favorite-longshot bias' state, 
which can be seen more 
clearly below.
The discrepancy between AR and EAR is   the largest at 
$t=t_{2}$ after $t=t_{1}$. After that, the discrepancy decreases 
monotonically as EAR increases faster than AR. The bets begin to be 
concentrated  on more popular horses. At $t=t_{3}$, AR and EAR coincide
again. The necessary condition of the market efficiency is
satisfied again. After $t_{3}$, the degree of the concentration
increases further, the discrepancy between AR and EAR is small
even at $t_{4}$.

\begin{figure}[t]
(a)
\begin{center}
\includegraphics[width=7.0cm,clip]{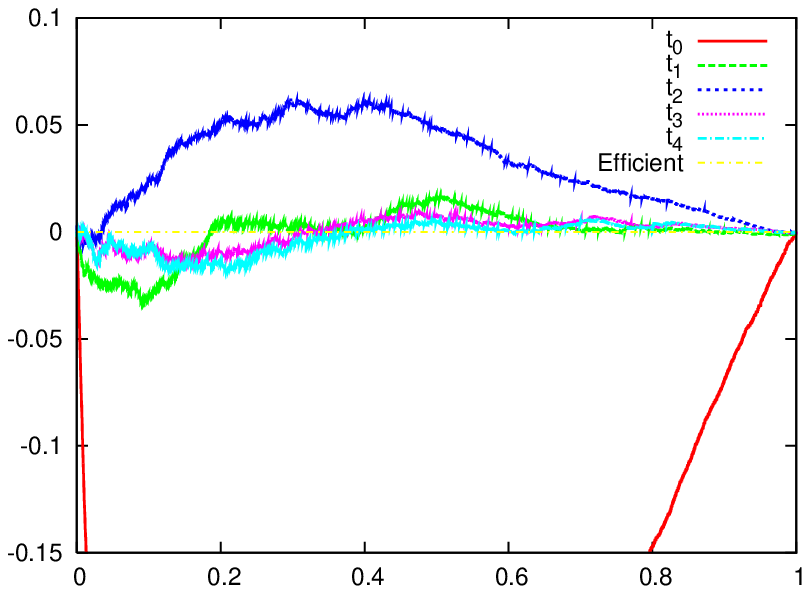}
\end{center}
(b)
\vspace*{-2.5cm}
\begin{center}
\includegraphics[width=11.0cm,clip]{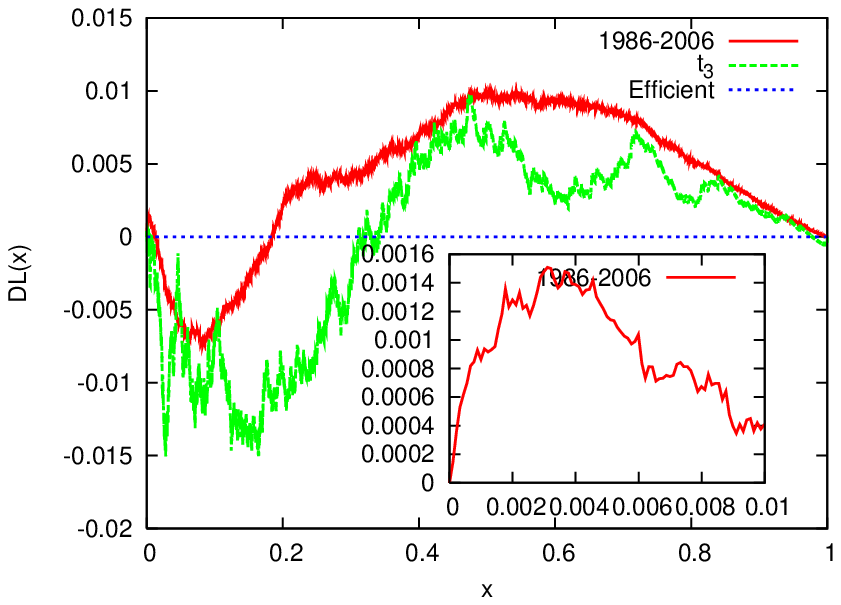}
\end{center}
\caption{(a) Plot of DL$(x)\equiv \mbox{L}(x)-\mbox{EL}(x)$ for $t=t_{i},i\in \{0,1,2,3,4\}$. 
(b) Plot of DL$(x)$ for $t_{3}$ and 
all the data of JRA races $(1986-2006)$.
Inset figure show the plot for all the data case for $0\le x \le 1\%$.}
\label{fig:DL}
\end{figure}

Figure \ref{fig:DL}a shows the discrepancies between 
$\mbox{L}(x)$ and $\mbox{EL}(x)$ at $t=t_{i},i\in \{0,1,2,3,4\}$.
On the y-axis, we show $\mbox{DL}(x)\equiv \mbox{L}(x)-\mbox{EL}(x)$.
As we have explained before, the sign of 
$\mbox{DL}'(x)=\frac{d \mbox{DL}(x)}{dx}$ tells us whether the horse at
rank $x$ is overestimated or underestimated. If $\mbox{DL}'(x)$ is zero,
the strength of the horse is properly estimated by the betters and the
racetrack betting market is efficient. On the other hand, if
$\mbox{DL}'(x)$ is positive (negative), the strength is underestimated 
(oversetimated).

At $t=t_{1}$, DL$(x)$ is close to the x-axis and we see that for small
$x\le 20\%$, 
$\mbox{DL}(x)<0$. 
This means that the market 
is almost efficient, but top 10\% popular horses are overestimated and 
next 10\% horses are underestimated. Bets are more accumulated on the
top 10\%
horses and their win bet fractions are larger  than their true winning
probabilities. On the other hand, there are less bets on next 10\% 
horses as compared to their true winning probabilities.
Remaining 80\% horses' strength are properly estimated, because DL$(x)$
 is close to the x-axis. 
At $t=t_{2}$, $\mbox{DL}(x)$ is positive for all $0 \le x \le 1$. 
From the figure, we see that  
the popular 30\% horses are underestimated as compared to their 
winning probabilities.  Remaining 70\% unpopular horses are overestimated.
The bets are
distributed among many horses, including many weak horses and 
an inefficient state is realized.
Following this, the graph of $\mbox{DL}(x)$ approaches the x-axis at $t=t_{3}$.
The coincidence between the two Lorenz curves is better than at $t=t_{1}$.
For small $x$ (popular horses), 
$\mbox{DL}(x)<0$ and the strong horses are overestimated.
For large $x$ (unpopular horses),  DL$(x)$ is almost on the x-axis.  
From the small discrepancy, we see that the top 20\% horses are 
overestimated. And next 30\% horses are underestimated.
However, the slope of DL$(x)$ in the two regions is small  and 
the market is almost efficient. This efficient state remains to be true even
at $t_{3}$.

Figure \ref{fig:DL}b shows DL$(x)$
 for the data of all JRA races (1986-2006). 
Contrary to the `favourite-longshot
bias', we see some complex behaviour.
Top 0.4\% horses are underestimated and next 10\% horses are
overestimated and so on.
However, the degree of the inefficiency
is very small.

\section{Voting model and scale invariance}

We consider a voting model for $N$ horses \cite{Mori}.
Betters vote for them one by one, and the result of each voting
is announced promptly. 
The time variable $t\in\{0,1,2\cdots,T\}$ 
counts the number of the votes.
The horses are classified into two categories $\mu\in \{0,1\}$, and we call
them binary horses.
There are $N_{\mu}$  horses in each category and 
$N_{0}+N_{1}=N$. 

We denote the number of votes of the $i$th horse $\mu\in \{0,1\}$
at time $t$ as $\{X^{\mu}_{i,t}\}_{i\in \{1,\cdots,N_{\mu}\}}$.
At $t=0$, $X_{i,t}^{\mu}$ takes the initial value
$X_{i,0}^{\mu}=s_{\mu}>0$. If the $i$th candidate $\mu$ gets a vote at $t$,
$X_{i,t}^{\mu}$ increases by one unit.
\[
X_{i,t+1}^{\mu}=X_{i,t}^{\mu}+1. 
\]
A better casts a vote for the 
$N$ candidates at a rate proportional to $X_{i,t}^{\mu}$. 
The probability $P_{i,t}^{\mu}$ that the $i$th 
candidate $\mu$ gets a vote at $t$ is
\begin{eqnarray}
P_{i,t}^{\mu}&=&\frac{X^{\mu}_{i,t}}{Z_{t}}, 
\\
Z_{t}&=&\sum_{\mu=0}^{1}
\sum_{i=1}^{N_{\mu}}X^{\mu}_{i,t}=N_{1}s_{1}+N_{0}s_{0}+t.
\end{eqnarray}

The problem of determining the 
probability of the $i$th candidate $\mu$ getting $n$ votes 
up to $T$ is equivalent to the 
famous P\'{o}lya's  urn problem \cite{Polya,Hisakado2}. 
The probability that the $i$th candidate $\mu$ gets $n$ votes up to $T$
is given by the beta binomial distribution
\begin{equation}
\mbox{Prob}(X_{i,T}^{\mu}-s_{\mu}=n)={}_{T}C_{n}\cdot 
\frac{(s_\mu)_{n}(Z_{0}-s_\mu)_{T-n}}{(Z_{0})_{T}}.
\end{equation}
$(a)_{n}$ is defined as $(a)_{n}=\frac{\Gamma(a+n)}{\Gamma(a)}$.

After infinite counts of voting, i.e. $T\to \infty$, 
the share of votes 
$x_{i}^{\mu}\equiv \lim_{T\to \infty}
\frac{X_{i,T}^{\mu}-s_{\mu}}{T}$ 
becomes the beta distributed random variable
beta$(s_{\mu},Z_{0}-s_{\mu})$ on $[0,1]$.
\begin{eqnarray}
p(x)&\equiv& \lim_{T\to \infty}
\mbox{Prob}(X_{i,T}^{\mu}-s_{\mu}=Tx)\cdot T \nonumber \\
&=&
\frac{x^{s_{\mu}-1}(1-x)^{Z_{0}-s_{\mu}-1}}{B(s_{\mu},Z_{0}-s_{\mu})}.
\end{eqnarray}

Next, we focus on the  thermodynamic limit $N_{0},N_{1} \to
\infty$ and $Z_{0}=N_{0}s_{0}+N_{1}s_{1}\to \infty$. 
The expectation value of $x_{i}^{\mu}$ is $<x_{i}^{\mu}>=p_{\mu}=
\frac{s_{\mu}}{Z_{0}}$. We introduce a variable $u_{i}^{\mu}\equiv
(Z_{0}-s_{\mu}-1)x_{i}^{\mu}$. 
The distribution function $p_{s_{\mu}}(u)$ in the  thermodynamic 
limit is given as
\begin{equation}
p_{s_{\mu}}(u)\equiv \lim_{Z_{0}\to \infty}p(x_{i}^{\mu}
=\frac{u}{Z_{0}-s_{\mu}-1})
=\frac{1}{\Gamma(s_{\mu})}e^{-u}u^{s_{\mu}-1}  \label{Eq:Gamma}.
\end{equation}
The share of votes, $u$, of a candidate $\mu$ 
follows a gamma distribution function with $s_{\mu}$.

After many counts of voting, $T\to \infty$, 
the two types of horses are distributed in the
space of $u$ according to the gamma distribution 
in the thermodynamic limit $Z_{0}\to \infty$.  If $s_{1}>s_{0}$,
 a candidate belonging to category $\mu=1$ has a higher 
probability of 
getting many votes than a candidate belonging to category $\mu=0$. 
Even the latter can obtain  
many votes. 
It is also possible that the former may 
get few votes. Thus, there is a mixing of the 
binary candidates.

The cumulative functions $x_{\mu}(w)$ is
given as
\begin{equation}
x_{\mu}(w)=\int_{0}^{w}p_{s_\mu}(u)du. 
\end{equation}
Using the incomplete gamma function of the first kind 
$\gamma(s,w)\equiv \int_{0}^{w}e^{-u}\cdot u^{s-1}du$,
 it is given as
\begin{equation}
x_{\mu}(w)=\frac{1}{\Gamma(s_\mu)}\cdot \gamma(s_\mu,w). \label{eq:ROC}
\end{equation}
Near the end point, $w \to 0$, in other words, in 
the small $u$ region, the incomplete gamma
function $\gamma(s_\mu,t)$  behaves  as
\begin{equation}
\gamma(s_\mu,w)\sim  w^{s_\mu}.
\end{equation}
As $x_{s_\mu}(w)\propto w^{s_\mu}$, the following relation
holds:
\begin{equation}
x_{1}\sim x_{0}^{\alpha}  \hspace*{0.4cm}
\mbox{with}\hspace*{0.4cm}
\alpha=\frac{s_1}{s_0}.      
\end{equation}
We see that a scale invariant behaviour appears 
in the mixing. 

Furthermore, in the limit $(s_1,s_0)\to (0,0)$ with 
fixed $\alpha=s_1/s_0$,
next relation holds \cite{Mori}.
\begin{equation}
x_{1}=x_{0}^{\alpha} \label{eq:scale}
\end{equation}
The scale-invariant relation holds over the entire
range $0\le x_{0},x_{1} \le 1$. This feature is remarkable from the
viewpoint of statistical physics. Usually, 
the power-law relation holds only in the tail.

We discuss the limit in the derivation of the exact scale
invariance. 
In the derivation
of the gamma distribution, we take the thermodynamic limit 
$Z_{0}=N_{1}s_{1}+N_{0}s_{0}\to \infty$. With the gamma distribution, 
$x_{1}=x_{0}^{\alpha}$ holds in the limit $\{s_{\mu}\}\to 0$.  
For (\ref{eq:scale}) to hold, these two limits, $Z_{0}\to
\infty$
and $\{s_{\mu}\}\to 0$, should go together. 
$\{s_{\mu}\}$ approaches zero more slowly than 
$\{N_{\mu}\}$ approaches infinity. We call the limit 
$Z_{0}\to \infty$ and $\{s_{\mu}\} \to 0$ with fixed 
$\alpha=s_{1}/s_{0}$
as the double scaling 
limit. 
If we take the limit $\{s_{\mu}\}\to 0$ without 
the limit $Z_{0}\to \infty$, the firstly chosen candidate gets all the 
remaining votes and there is no
mixing of the binary candidates. The double scaling limit
is  crucial to the emergence of the exact scale invariance.

\section{Exact scale invariant gradation pattern}

The voting problem reduces to  a random ball removing problem with the  
relative probability $s_\mu$ in the double scaling limit \cite{Mori}.
 Figure 11 shows the gradation pattern made by the algorithm of the random ball
 removing problem. We prepare $N_{1}$ red balls and $N_{0}$ blue balls. 
From them, we 
take one ball at a time and do not return it. The probability that 
a red (blue) ball is chosen is proportional to $s_{1} (s_{0})$.
We repeat the procedure up to when there remains no ball. We get 
a sequence of $N=N_{1}+N_{0}$ balls. In the limit 
$N_{1},N_{0}\to \infty$, the exact scale invariance between $x_{1}$ and 
$x_{0}$ holds. In the figure,
 we change the ratio $\alpha=\frac{s_{1}}{s_{0}}$
 from 1 (bottom) to 100 (top). Near the bottom,
 two types of balls are mixed. Near the top, phase separation 
occurs.

\begin{figure}
\begin{center}
\includegraphics[width=8cm,clip]{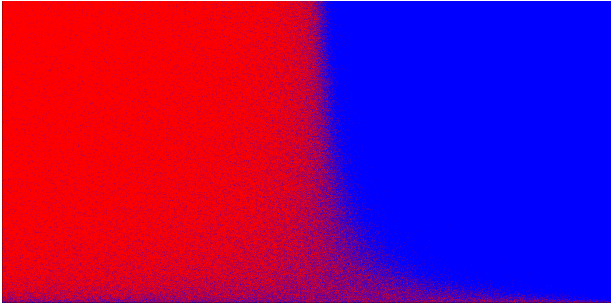}
\end{center}
\caption{Exact scale invariant gradation pattern with varying $\alpha$
from $\alpha=1$ (bottom) to $\alpha=100$ (top).}
\label{fig:EXACT}
\end{figure}

\section{Summary}

In this paper, we study the time series of win bet fraction of the 2008
JRA races. We see that the betting process induces scale invariance 
between the cumulative functions of the winning horses and that of the 
losing horses.
We also see the emergence of the market's efficiency after many 
rounds of betting. However, the convergence to the efficient state is not 
monotonic. The dynamics of the distribution of the votes among the horses
is complex. At first, the votes accumulate to  small number of horses and 
then they are distributed among many horses, including weak horses.
At this time, the strong horses are underestimated. 
Afterwards, votes begin to be
concentrated on more popular horses, but the ranking of the winning
horses does not change 
so much. AR does not  change much and only EAR increases, and
$\mbox{AR}=\mbox{EAR}$ finally holds.

With regard to the scale invariance, 
we explain the mechanism based on a simple
voting model. The model shows exact scale invariance in the 
double scaling limit. In addition, the voting model in the limit is 
equivalent to a random ball removing problem. Using the equivalence,
 we show how to make an exact gradation pattern of mixed binary objects.

\section{Acknowledgment}

This research was partially supported by the Ministry of Education,
Science, Sports and Culture, Grant-in-Aid for 
Challenging Exploratory Research, 21654054, 2009.


\begin{thebibliography}{99}

\bibitem{Ziemba} D.B.Hausch, V.SY.Lo and T.Ziemba, 
{\it Efficiency of Racetrack Betting Markets 2008 Edition}, World
	Scientific, Singapole.

\bibitem{Griffith} R.M.Griffith, Amer.J.Psychol {\bf 62}290(1949). 

\bibitem{Ziemba2} W.T.Ziemba and D.B.Hausch, {\it Betting at the Racetrack
},Dr Z Investment Inc, San Luis Obispo, CA(1987).

\bibitem{Park} K.Park and E.Dommany, Europhys.Lett. {\bf 53},419(2001).

\bibitem{Ichinomiya} T.Ichinomiya, Physica {\bf A368},207(2006).

\bibitem{Mori} S.Mori and M.Hisakado, 
{\it Exact Scale Invariance in
	Mixing of Binary Candidates in Voting Model}, preprint arXiv:0806.0185.

\bibitem{Hisakado} M.Hisakado and S.Mori, {\it Phase Transition and
	Information Cascade in a Voting Model}, preprint arXiv:0907.4818.

\bibitem{Votes} More precisely, we calculate $\hat{x}^{r}_{i,k}$ before 
removing the cancelled horses from data. Up to the cancellation, these
do not sum up to one. After cancellation, the odds and the win bet
	fraction become zero. These sum up to one. 

\bibitem{EL}
If cancellation occurs in the betting process, the sum of $x_{\alpha}$ is
not $R$ before cancellation. In this case, it is necessary to divide
by $\sum_{\alpha=1}^{H}x_{\alpha}$


\bibitem{Polya} G.P\'{o}lya, Ann.Inst.Henri Poincar\'{e} {\bf 1},117(1931). 


\bibitem{Hisakado2} M.Hisakado, K.Kitsukawa and S.Mori, J.Phys. {\bf
	A39},15365(2006).

\bibitem{Enleman} B.Enleman, E.Hayden and D.Tasche,
{\it Testing Rating Accuracy}, www.risk.net (2003).


\end{thebibliography}
\end{document}